\pdfoutput=1
\documentclass[
 reprint,
 superscriptaddress,
 amsmath,amssymb,
 aps,
 prd
]{revtex4-2}
\usepackage[T2A, T1]{fontenc}
\usepackage[final,breaklinks]{hyperref}
\usepackage[russian,german,french,latin,english]{babel}
\usepackage{CJKutf8}
\usepackage{graphicx}
\usepackage{dcolumn}
\usepackage{bm}
\usepackage{xcolor,tikz-cd,mleftright,mathdots,mathtools}
\usepackage[smalltableaux]{ytableau}
\usepackage[final]{hyperref}
\usepackage{cleveref,orcidlink}
\let\detold\det
\renewcommand\det{\detold\nolimits}
\newcommand{\sst}[1]{{\scriptscriptstyle (#1)}}
\hypersetup {
    pdftitle = {Duality Anomalies in Linearized Gravity},
    pdfauthor = {Leron Borsten, Michael J. Duff, Dimitri Kanarakis, Hyungrok Kim},
    pdfsubject = {hep-th, gr-qc, math-ph},
    pdflang = {en-US},
    pdfkeywords = {duality anomaly, linearized gravity, dual graviton, Batalin–Vilkovisky formalism, Ray–Singer torsion},
}

\begin{document}

\title{Duality Anomalies in Linearized Gravity}

\author{Leron  Borsten}
 \email{l.borsten@herts.ac.uk}
\affiliation{%
Department of Physics, Astronomy and Mathematics\\University of Hertfordshire, Hatfield, Herts.\ Al10 9AB, United Kingdom
}%
\affiliation
 {Blackett Laboratory, Imperial College London, London SW7 2AZ, United Kingdom
}
\author{Michael J.~Duff}
 \email{m.j.duff@imperial.ac.uk}
\affiliation
 {Blackett Laboratory, Imperial College London, London SW7 2AZ, United Kingdom
}

\author{Dimitri  Kanakaris}%
 \email{d.kanakaris-decavel@herts.ac.uk}
\affiliation{%
Department of Physics, Astronomy and Mathematics\\University of Hertfordshire, Hatfield, Herts.\ Al10 9AB, United Kingdom
}%

\author{Hyungrok Kim (\begin{CJK*}{UTF8}{bsmi}金炯錄\end{CJK*})}
 \email{h.kim2@herts.ac.uk}
\affiliation{%
Department of Physics, Astronomy and Mathematics\\University of Hertfordshire, Hatfield, Herts.\ Al10 9AB, United Kingdom
}

\begin{abstract}
Classical linearized gravity admits a dual formulation in terms of a higher-rank tensor field. However, proposing a prescription for the instanton sectors of linearized gravity and its dual, we show  that they may be \emph{quantum inequivalent}.   The duality anomaly is obtained by resolving  the dual graviton theories  into vector-valued \(p\)-form electrodynamics and is controlled by   three topological invariants: the  Reidemeister torsion,  Ray--Singer torsion and twisted Euler characteristic.  Under the proposed instanton prescription the duality anomaly vanishes for an odd number of spacetime dimensions as a consequence of the celebrated Cheeger--M\"uller theorem. In the presence of a gravitational $\theta$-term, the partition function is a modular form  in direct analogy to Abelian S-duality for Maxwell theory. 
\end{abstract}

\maketitle

\section{\label{sec:level1}Introduction}
It sometimes happens that a  theory admits a \emph{dual} description that, although ostensibly distinct, is in fact physically equivalent.    The most relevant example here  is  Montonen--Olive/S-duality in (supersymmetric) quantum field theory \cite{Montonen:1977sn, Witten:1978mh, Osborn:1979tq, Duff:1994zt}, but dualities have played pivotal role in the development of modern physics across diverse domains: bosonization \footnote{Bosonization can be viewed as an instance of S-duality. Indeed, the S-duality of the Thirring and sine-Gordon models  is the direct field theory analogy; the sine-Gordon model is the S-dual  bosonization of the  Thirring model \cite{Coleman:1974bu}.} in  Luttinger liquids and Majorana zero modes \cite{Luttinger:1963zz, Mattis:1964wp, Fidkowski:2011rf}; Kramers--Wannier duality  in the Ising model \cite{Kramers:1941zz}; particle-vortex duality in the fractional quantum Hall eﬀect \cite{Shapere:1988zv, Fisher:1989dnp, Rey:1989jg, Lee:1991jt, Burgess:2000kj};  S-duality in symmetry protected topological phases \cite{Metlitski:2015yqa}; U-duality in supergravity and string/M-theory \cite{Cremmer:1978ds, Cremmer:1979up,Julia:1981wc,Cremmer:1997ct,Hull:1994ys},  to name but a few. Often such dualities  shed new light on otherwise inaccessible facets of the phenomenology \footnote{These duality insights have been to numerous to do justice to here, but to convey the broad scope let us mention just two seemingly disparate examples: Kramers--Wannier duality  in locating the  critical point in the two-dimensional Ising model \cite{Kramers:1941zz} and U-duality in the microscopic derivation of the Bekenstein--Hawking entropy of black holes in string theory \cite{Strominger:1996sh}.} and  may reveal deep links to mathematics, such as mirror symmetry and the geometric Langlands correspondence \cite{Kapustin:2006pk}.

An early example is the classical electric--magnetic duality of $d=3+1$ Maxwell theory,  a  precursor to  Montonen--Olive S-duality.  Electric--magnetic duality  requires magnetic monopoles which, as Dirac argued, would imply  electric charge quantization \cite{Dirac:1948um}. More generally, Abelian \(p\)-form gauge theories \cite{Teitelboim:1985ya} in \(d\) spacetime dimensions enjoy a classical electric--magnetic duality \cite{Henneaux:1986ht} in which the \(p\)-form gauge potential \(A\) is dualized to a \((d-2-p)\)-form potential \(\tilde A\) via the relation $
    \mathrm dA=\star\mathrm d\tilde A,$
where \(\star\) is the Hodge dual.  Such electric--magnetic duality is a pervasive phenomenon in classical gauge theory, at least for free theories, cf.~\cite{deMedeiros:2002qpr, Bansal:2021bis, Avetisyan:2021heg}. However, a  classical  electric--magnetic duality  may be anomalous  at the quantum level   \cite{Duff:1980qv,Schwarz:1984wk,Witten:1995gf,Olive:2000yy,Donnelly:2016mlc}. Rather than thinking of this as a failure, there are at least two opportunities presented by such an obstruction. First, constraining the theory to achieve   anomaly freedom (covariance) frequently  entails  important insights \footnote{Again, there are too many instances to catalog, and we merely mention three examples indicating the diverse applications:  constraining the bound state spectrum of confining gauge theories via  ’t Hooft anomaly matching conditions \cite{tHooft:1979rat}; critical dimensions \cite{Polyakov:1981rd} and Einstein's equations \cite{Callan:1985ia,Fradkin:1984pq} from the  vanishing of Weyl anomaly \cite{Capper:1974ic,Duff:1977ay} on the string world-sheet;  the identification of five consistent superstring theories via the Green--Schwarz mechanism \cite{Green:1984sg}.}, as in Dirac's charge quantization condition. Second, the  dual description may allow for a more favorable  quantization. Historically, both takes have proved effective.

Given these successes,  it is natural to seek generalizations to the more challenging context of gravitational dualities. 
 In particular, there exists  a dual formulation \cite{Curtright:1980yk,Hull:2000zn, Hull:2000rr, West:2001as, Hull:2001iu, Hull:2023iny,Hull:2024qpy} of linearized gravity  propagating on a background manifold $(M,g)$. The perturbative graviton, \(h_{\mu\nu}\), is replaced by a vector-valued \((d-3)\)-form, \(\tilde h_{\mu_1\dotso\mu_{d-3}\nu}\), with corresponding symmetry constraints \footnote{There also exists a ``double-dual'' graviton, where one dualizes both indices rather than a single index; in four dimensions, this is however a mere algebraic relabeling \cite{Henneaux:2019zod}.}. Going beyond the free theory, in \cite{West:2001as} an action including  the full  dual graviton  \(\tilde g_{\mu_1\dotso\mu_{d-3}\nu}\)  and  graviton $g_{\mu\nu}$ together is introduced and shown to yield a theory  equivalent to Einstein--Hilbert gravity \footnote{There are obstructions to going beyond the linear case \cite{Bekaert:2002uh,Monteiro:2023dev}. These are evaded in \cite{West:2001as} by the fact that the local dual graviton equation of motion includes the graviton itself. In this case the dual graviton appears as an auxiliary field, which may be eliminated to recover the Einstein--Hilbert action. Integrating out, instead, the graviton would yield a non-local action, violating one of the assumptions entering \cite{Bekaert:2002uh}.}, establishing directly the equivalence of the linearized dual graviton theories.  Although standing alone, such  dual formulations of gravity also inevitably   arise  in    manifesting M-theory dualities \cite{Hull:2000zn, Hull:2000ih, Hull:2000rr, West:2001as, Henneaux:2016opm, Hohm:2018qhd, Tumanov:2017whf, Glennon:2020uov, Hull:2022vlv,Boulanger:2024lwk} and   in gravitational  generalized symmetries \cite{Hinterbichler:2022agn,Benedetti:2021lxj,Benedetti:2023ipt,Benedetti:2022zbb,Gomez-Fayren:2023qly,Hull:2024bcl,Hull:2024ism}. 
 
 It is natural then to  ask whether the duality anomaly persists. We address this question here. The answer rests on elegant relations among   topological invariants, notably the    Cheeger--Müller theorem \cite{zbMATH03558610,zbMATH03640582,zbMATH03613851,muellerthesis}, and subtle points regarding the nature of  instantons in linearized gravity. Our analysis relies   crucially on resolving the symmetry constraints of   the linearized (dual) graviton,  using the Batalin--Vilkovisky (BV) formalism  \cite{Batalin:1977pb,1981PhLB..102...27B,1983PhRvD..28.2567B,BATALIN1984106,Batalin:1985qj}, to rearticulate it  as a ${\rm T}^*M$-valued ($d-3$)-form field. This implies that the positive-energy mode contributions to  the dual partition functions combine into the  Ray--Singer analytic torsion, as witnessed for dual $p$-form Maxwell theories \cite{Schwarz:1984wk}.  In the latter case, the zero modes and instantons  conspire to cancel the anomaly in odd dimensions \cite{Donnelly:2016mlc}. However,  the corresponding treatment in the gravitational  setting  is less clear since the linearization of the ``instantons''  is not obvious \footnote{This is analogous to recovering the ``linearized instanton moduli space'' \(\operatorname H^2(M;\mathbb Z)\otimes\mathfrak g\) from the moduli space of \(\mathfrak g\)-valued Yang--Mills instantons on a spacetime \(M\).}.
Here, we propose a prescription for the instantons  as elements of ${\rm H}^{p+1}(M; \mathbb{Z})$ and compute the duality anomaly for linearized gravity on a flat background metric. We find that the anomaly is
\begin{equation}\label{eq:grav-anomaly}
    Z_\text{grav}/\tilde Z_\text{grav} =          (\kappa/\tilde \kappa)^{\tfrac12\chi(M;\mathrm T^*M)} ,
\end{equation}
where $\kappa$  (\(\tilde \kappa\)) is the  (dual) linearized gravity coupling constant and  \(\chi(M;\mathrm T^*M)\) is the Euler characteristic for the cohomology of \(M\) twisted with respect to the local system given by \(\mathrm T^*M\) \footnote{In particular, when \(M\) is parallelizable, then \(\chi(M;\mathrm T^*M)=\chi(M)d\).}. The purely topological characterization of the anomaly and its absence  in odd dimensions a~posteriori justify our instanton prescription. Moreover,  it suggests a number of entailments regarding the definition of the partition function for gravity and dualities in string/M-theory  that we expand on in the conclusions.

\section{The Cheeger--Müller theorem}
Suppose that we are given a closed oriented connected Riemannian manifold \((M,g)\) and a vector bundle \(E\twoheadrightarrow M\) equipped with a flat connection and a compatible metric. 

\subsection{Ray--Singer analytic torsion.}
The space of \(E\)-valued differential forms \(\Omega^\bullet(M;E)\) has   canonical covariant (since $E$ is flat) differential
$
    \mathrm d\colon \Omega^\bullet(M;E)\to \Omega^{\bullet+1}(M;E)
$.
The metrics on $M$ and $E$ induce an adjoint 
$
    \mathrm d^\dagger\colon \Omega^\bullet(M;E)\to \Omega^{\bullet-1}(M;E)
$,
in terms of which one can define the Laplace--de~Rham operator $\Delta_p^E\colon \Omega^p(M;E) \to \Omega^p(M;E)$ as
$
    \Delta_p^E = \mathrm d\mathrm d^\dagger + \mathrm d^\dagger\mathrm d
$.
Since \(M\) is compact, \(\Delta_p^E\) is positive-semidefinite with pure point spectrum $\{\lambda_n\}$. Ray and Singer defined the zeta-function-regularized determinant restricted to the strictly positive spectrum,
\begin{equation}
    \det'(\Delta_p^E) = \exp\mleft(-(\zeta^E_k)'(0)\mright),
\end{equation}
via the analytic continuation of
$
    \zeta_k^E(s) = \sum_{\lambda_n>0}\lambda^{-s}. 
$
The Ray--Singer analytic torsion \cite{zbMATH03379873,zbMATH03421163,zbMATH03429853} is then defined as
\begin{equation}
    \tau_\mathrm{RS}(M;E) = \prod_{k=0}^d\det'(\Delta^E_k)^{-(-1)^kk/2}.
\end{equation}

\subsection{Reidemeister torsion.}
Consider the \(E\)-valued cohomology \(\operatorname H^k(M;E)\) (with integer coefficients). This is a finitely generated Abelian group, such that we may decompose it (non-canonically) as a direct sum of a torsion-free part \(\operatorname{Free}(\operatorname H^k(M;E))\) and the torsion part \(\operatorname{Tor}(\operatorname H^k(M;E))\),
\begin{equation}
    \operatorname H^k(M;E) \cong \operatorname{Free}(\operatorname H^k(M;E))\oplus\operatorname{Tor}(\operatorname H^k(M;E)).
\end{equation}
 Picking a topological basis \(\{w_i\}\) on \(\operatorname{Free}(\operatorname H^k(M;E))\), we may represent the metric  as a \(b_k^E\times b_k^E\) matrix $[\Gamma_k]_{ij} = [M]\frown ([w_i]\smile [w_j])$  \footnote{Here, \(\smile\) is the cup product between cohomology classes, and \(\frown\) is the cap product between homology and cohomology classes \cite{zbMATH02103273}.}, where \(b_k^E\) is the rank (\(k\)th Betti number) of the free \(\mathbb Z\)-module \(\operatorname{Free}(\operatorname H^k(M;E))\).
The Reidemeister torsion \cite{zbMATH03017467,zbMATH03018437,zbMATH03025073} is the quantity
\begin{equation}
    \tau_\mathrm{Reid}(M;E) 
    = 
    \prod_{k=0}^d\det \Gamma_k^{(-1)^k/2}|\operatorname{Tor}(\operatorname H^k)|^{(-1)^{k+1}},
\end{equation}
where \(|G|\) denotes the order (cardinality) of a finite group \(G\).
(For the torsion subgroup factor, see \cite[App.~E]{Metlitski:2015yqa}.)
It can be shown that it does not depend on the arbitrary choice of topological basis, so that it is an invariant of the topological manifold \(M\) and the flat vector bundle \(E\).

The \emph{Cheeger--Müller theorem} \cite{zbMATH03558610,zbMATH03640582,zbMATH03613851,muellerthesis} states that, when \(\operatorname{Tor}(\operatorname H^k(M;E))=0\), 
$    \tau_\mathrm{RS}(M;E) = \tau_\mathrm{Reid}(M;E), 
$ as conjectured by Ray and Singer. 

\section{Dualities for \(p\)-form electrodynamics}
In the partition function of higher gauge theories \cite{Borsten:2024gox} (such as \(p\)-form electrodynamics), the ghosts, ghosts-for-ghosts, and so on are important \cite{Siegel:1980jj}.  To this end, we employ the BV formalism.

The BV action for an Abelian \(p\)-form gauge  potential \(A\) (valued in the trivial line bundle $E=M\times \mathbb{R}$) is 
\begin{equation}
\begin{split}
    S &= \frac1q \int \mathrm dA\wedge\star\mathrm dA
    + A^+\wedge\mathrm dc^{\sst{0}}
    + c_{\sst{-1}}^+\wedge\mathrm dc^{\sst{-1}}\dotsb\\
    &~~~~~~~~~~ \dots
    + c_{\sst{2-p}}^+\wedge\mathrm dc^{\sst{1-p}},
        \end{split}
\end{equation}
where $q$ is the coupling constant. In addition to $A$, this action involves the tower of \((p-1+i)\)-form field ghosts \(c^{\sst{i}}\),  with ghost number \(1-i\), 
as well as the corresponding antifields \(A^+,c^{+}_{\sst 0},\dotsc,c^{+}_{\sst{1-p}}\). Gauge fixing involves \cite[§4.4]{Jurco:2018sby} the introduction of a large number of trivial pairs of Nakanishi--Lautrup fields and antighosts, 
$(b_{\sst{i,j}},\bar c_{\sst{i,j}})$, where $i\in\{1-p,\dotsc,0\},\; j\in\{i-1,i+1,\dotsc,-i-3,-i-1\}$, and their corresponding antifields. Here, \(\bar c^{\sst{i,j}}\) is a \((p-1+i)\)-form field of ghost number \(j\) and \(b^{\sst{i,j}}\) is an auxiliary  \((p-1-i)\)-form field of ghost number \(j+1\). After gauge fixing in (the \(p\)-form analogue of) Feynman gauge and integrating out the auxiliary Nakanishi--Lautrup fields \(b^{\sst{i,j}}\), we are left with
\begin{equation}\label{eq:p-formBV}
\begin{split}
    S &= \frac{1}{ q}\int \tfrac12A\wedge\star \Delta A +\sum_{i={1-p}}^0\bar c^{\sst{i,i-1}} \wedge\star\Delta c^{\sst{i}}\\
    &~~~~~ +
    \tfrac12\sum_{i=1-p}^0\sum_{\substack{j=i+1\\i\not\equiv j\pmod2}}^{-i-1}\bar c^{\sst{i,-j}}\wedge\star\Delta\bar c^{\sst{i,j}},
    \end{split}
\end{equation}
where \(\Delta=\mathrm d\mathrm d^\dagger+\mathrm d^\dagger\mathrm d\) is the Laplace--de~Rham operator. 

The partition function $Z_p = \int\mathrm DA\,\mathrm Dc\,\exp{S}$, where we have denoted the measure for the ghost tower by $\mathrm Dc$, splits into three contributions: (i) the positive eigenvalues of \(\Delta\), which involve the zeta-regularized determinant \(\det'\Delta\); 
(ii) the zero modes of \(\Delta\), which involve \(\operatorname H^k(M;\mathbb Z)\) for \(k\le p\); (iii)
a sum over the possible \(\operatorname U(1)\) \((p-1)\)-gerbes \cite{Borsten:2024gox,Alfonsi:2023pps} on \(M\), which involves \(\operatorname H^{p+1}(M;\mathbb Z)\).
In evaluating the ratio of the dual partition functions
$    Z_p/\tilde Z_{d-p-2}$ that characterizes the duality anomaly of \(p\)-form electrodynamics, all three contributions come into play. The first contribution organizes itself into the Ray--Singer torsion \cite{Schwarz:1984wk}; the second and third contributions organize themselves into the Reidemeister torsion with some additional factors \cite{Donnelly:2016mlc}. When \(d\) is odd, the two contributions precisely cancel each other out due to the Cheeger--Müller theorem; when \(d\) is even, one finds \cite[(1.1)]{Donnelly:2016mlc}
\begin{equation}
    Z_p/\tilde Z_{d-p-2}
    =(q/\tilde q)^{\tfrac12(-1)^{p+1}\chi(M)},
\end{equation}
where \(\chi(M)\) is the Euler characteristic of \(M\) and \(\tilde q = 2\pi/ q\) is dual  the coupling constant.

\section{Linearized (dual) gravity}\label{sec:main}

In this section, \((M,g)\) is a closed oriented Riemannian manifold. We further assume   \(g\) is flat so that  \(\mathrm T^*M\) is flat and we may employ the Cheeger--Müller theorem.

\subsection{${\rm T}^*M$-valued \(p\)-form resolution of the  graviton}
 Consider linearized gravity (with no cosmological constant) atop a background Riemannian manifold \((M,g)\). In terms of a perturbative metric
$
    \sqrt\kappa h_{\mu\nu} = g^\mathrm{dynamical}_{\mu\nu}-g_{\mu\nu}
$,
where \(g^\mathrm{dynamical}\) is the dynamical metric and \(\kappa=8\pi G\) is the gravitational constant, the massless BV Fierz--Pauli action for linearized gravity is simply
\begin{equation}
	S_\mathrm{FP}
	= \frac{1}{\kappa^2}\int \left(\operatorname{vol}_g\,\mathcal{L}_{\mathrm{FP}}[h]\right)
		+ h^{+\mu\nu}\nabla_\mu X_\nu,
\end{equation}
where $\mathcal{L}_{\mathrm{FP}}[h]$ is the familiar massless Fierz--Pauli Lagrangian density, $\kappa^2=16 \pi G_{\mathrm{N}}^{(d)}c^{-3}$, \(\operatorname{vol}_g=\mathrm d^dx\,\sqrt{|\det g|}\) is the volume form,  \(h = h_{\mu\nu}g^{\mu\nu}\) is the trace,  \(h^{+\mu\nu}\) is the graviton antifield tensor density, and  \(X_\mu\) is the diffeomorphism ghost.

Harmonic gauge,
$    \partial_\mu h_\nu{}^\mu=\tfrac12\partial_\nu h
$, reduces the action to
\begin{equation}
    S_\mathrm{FP} = \frac{1}{\kappa^2} \int \operatorname{vol}_g\,\left(\tfrac14h^{\mu\nu} \Delta h_{\mu\nu}
    + \bar X_\mu \Delta X^\mu\right),
\end{equation}
where  \(\bar X_\mu\) is the diffeomorphism antighost and \(\Delta\)  is the Laplacian with respect to \(g\).

At this point, one could  express the positive-energy-mode contribution to the partition function in terms of the determinant of the Laplacian on symmetric rank-two tensor fields,
$    Z_\mathrm{osc} = (\det'(\Delta_0^{\mathrm TM\odot_M\mathrm TM}))^{-1/2} \det'(\Delta_1)
$ \cite{Christensen:1979iy}. However, we can short-circuit much of this by the following trick: similar to the generalized metric in generalized geometry (see e.g.\ \cite{Cavalcanti:2011wu,Zabzine:2006uz}), we may add an extra sector corresponding to a two-form (Kalb--Ramond-type) field \(B_{\mu\nu}\) and consider
\begin{equation}\label{eq:generalised_metric}
    H_{\mu\nu} = \frac1{\sqrt2}(h_{\mu\nu}+B_{\mu\nu}),
\end{equation}
which is a two-tensor without any symmetry properties. We may regard
$H_\nu = H_{\mu\nu}\,\mathrm dx^\mu$
as a \(\mathrm T^*M\)-valued one-form \footnote{Indeed, this is equivalent to starting from the frame-bundle picture and linearizing. The addition symmetries and ghosts that arise here correspond, in this picture, to the  local Lorentz ghosts.} so that  
\begin{equation}\label{eq:fierz-pauli-extended}
    S_{\text{FP},\text{gf}} = \frac{1}{\kappa^2}\left(S_1^{\mathrm{T}^*M} - S_2 + \bar S_1 - S_0\right)
\end{equation}
where
\begin{subequations}\label{eq:fierz-pauli-extended-detail}
\begin{equation}\label{eq:fierz-pauli-extended-detail1}
\begin{split}
    S_1^{\mathrm{T}^*M} &= \int\tfrac12H^\mu\wedge\star\Delta H_\mu
    + \bar X^\mu \wedge\star\Delta X_\mu \\
    \end{split}
\end{equation}
\begin{equation}\label{eq:fierz-pauli-extended-detail2}
\begin{split}
    S^{M\times\mathbb R}_2 &= \int\tfrac12B\wedge\star\Delta B + \sum_{i=-1}^{0}\bar c^{\sst{i,i-1}}\wedge\star\Delta c^{\sst{i}}\\
  & ~~~~~~~+ \tfrac12\bar c^{\sst{-1,0}}\wedge\star\Delta\bar c^{\sst{-1,0}} \\
  \end{split}
\end{equation}
\begin{equation}\label{eq:fierz-pauli-extended-detail3}
\begin{split}
    \bar S^{M\times\mathbb R}_1 &= \int \bar c^{\sst{0,-1}}\wedge\star\Delta c^{\sst{0}} + \bar c^{\sst{-1,-2}}\wedge\star\Delta c^{\sst{-1}} \\
  &
~~~~~~+ \tfrac12\bar c^{\sst{-1,0}}\wedge\star\Delta\bar c^{\sst{-1,0}} + \tfrac12\phi\wedge\star\Delta\phi\\
\end{split}
\end{equation}
\begin{equation}\label{eq:fierz-pauli-extended-detail4}
\begin{split}
     S^{M\times\mathbb R}_0 &= \int \tfrac12\phi\wedge\star\Delta\phi.
\end{split}
\end{equation}
\end{subequations}
Here
\(S^E_p\) is the gauge-fixed BV action for an Abelian $p$-form valued in the flat vector bundle $E$, cf.~\eqref{eq:p-formBV}.  
The  \(\bar S^{M\times\mathbb R}_1\) term may be thought of as two copies of Maxwell theory with wrong statistics, that is, a  pair of anticommuting vector fields \(\bar c^{\sst{0,-1}}\) and \(c^{\sst{0}}\) together with their requisite commuting scalar (anti)ghosts \(\bar c^{\sst{-1,-2}},c^{\sst{-1}},\bar c^{\sst{-1,0}},\phi\).

In terms of the degree-of-freedom count, the identity \eqref{eq:fierz-pauli-extended}  may be expressed as $\tfrac12d(d-3) =\left(d\binom d1 -2 d\binom d0 \right)-\left(\binom d2 -2 \binom d1 +3 \binom d0\right)-2(\binom d1-2\binom d0)-1$.

We therefore may write the partition function \(Z_\text{grav}\) for linearized gravity as
\begin{equation}\label{eq:partition-function-relation}
    Z_\mathrm{grav} = \frac{Z_1^{\mathrm T^*M} (\bar Z^{M\times\mathbb R}_1)^2}{Z^{M\times\mathbb R}_2Z^{M\times\mathbb R}_0} = \frac{Z_1^{\mathrm T^*M}}{Z^{M\times\mathbb R}_2(Z^{M\times\mathbb R}_1)^2Z^{M\times\mathbb R}_0},
\end{equation}
where we have used the fact that reversing the statistics inverts the partition function, $\bar Z_p^E = 1/{Z_p^E}$.

\subsection{${\rm T}^*M$-valued \(p\)-form resolution of the dual  graviton}
The dual graviton  \(\tilde h_{\mu\nu_1\dotso\nu_{d-3}}\) in \cite{Hull:2000zn,West:2001as,Hull:2001iu}   transforms in the \(\operatorname{GL}(d)\)-representation given by the Young diagrams
\begin{equation}
\overbrace{
\begin{ytableau}
{} & {} & \none[\scriptstyle\cdots] & {} \\
{} \\
\end{ytableau}}^{d-3}
=
\left(\ydiagram{1}\otimes
\overbrace{\begin{ytableau}
{} & {} & \none[\scriptstyle\cdots] & {} 
\end{ytableau}}^{\smash{d-3}}\right)
-
\overbrace{
\begin{ytableau}
{} & {} & \none[\scriptstyle\cdots] & {} & {}
\end{ytableau}}^{d-2}.
\end{equation}
That is, 
$ \tilde h_{\mu\nu_1\dotso\nu_{d-3}}=\tilde h_{\mu[\nu_1\dotso\nu_{d-3}]}$ and 
   $ \tilde h_{[\mu\nu_1\dotso\nu_{d-3}]}= 0.$
   For convenience, let us call such tensor a \([1,d-3]\)-tensor.

The field \(\tilde h\) transforms under a gauge transformation valued in a \(\mathrm T^*M\)-valued \((d-4)\)-form as 
\begin{equation}
    \delta\tilde h_{\mu\nu_1\dotso\nu_{d-3}}
    = \tilde X^{\sst{0}}_{\mu[\nu_1\dotso\nu_{d-4},\nu_{d-3}]}
    -
    \tilde X^{\sst{0}}_{[\mu\nu_1\dotso\nu_{d-4},\nu_{d-3}]}.
\end{equation}
When \(d=4\), then \(\tilde X^{\sst{0}}_\mu\) is merely \(\mathrm T^*M\)-valued 0-form, and this reduces to the standard linearized diffeomorphism
$    \delta\tilde h_{\mu\nu} =\tilde X^{\sst{0}}_{(\mu,\nu)},
$ while for \(d>4\),  \(\tilde X^{\sst{0}}_{\mu\nu_1\dotso\nu_{d-4}}\) decomposes into the irreducible \((d-3)\)-form and  \([1,d-4]\)-tensor  \(\operatorname{GL}(d)\)-representations, cf.\ \cite[(3.27)]{Hull:2001iu} and \cite[(2.1)]{Hull:2023iny}.
In formulating the complete BV action, it is convenient for us not to make this decomposition, however.

Additionally there is a tower of \(\mathrm T^*M\)-valued \((d-4-i)\)-form ghosts,  $i\in\{0,\dotsc,d-4\}$, for higher-order  gauge-for-gauge symmetries \cite{Aulakh:1986cb,Labastida:1986gy} given by  
\begin{align}
    \delta\tilde X^{\sst{-i}}_{\mu\nu_1\dotso\nu_{d-4-i}} &= \tilde X^{\sst{-i-1}}_{\mu[\nu_1\dotso\nu_{d-5-i},\nu_{d-4-i}]}.
\end{align}
Thus, the BV action is
\begin{equation}
\begin{split}
    \tilde S &= \tilde S_{\widetilde{\text{FP}}} +\frac{1}{\tilde \kappa^2}\left( \int   h^{+\mu\nu_1\dotso\nu_{d-3}}\tilde X^{\sst{0}}_{\mu[\nu_1\dotso\nu_{d-4},\nu_{d-3}]}\right.\\
    &\left. +\sum_{i=5-d}^{0}
    \tilde X_{\sst{i}}^{+\mu\nu_1\dotso\nu_{d-4+i}}\tilde X^{\sst{i-1}}_{\mu[\nu_1\dotso\nu_{d-5+i},\nu_{d-4+i}]}\right),
    \end{split}
\end{equation}
where \(\tilde S_{\widetilde{\text{FP}}}\) is the ordinary action for the dual graviton \cite{Curtright:1980yk,Aulakh:1986cb,Labastida:1986gy} and $\tilde \kappa = 2\pi / \kappa$.

Applying  a suitable gauge choice \cite{Aulakh:1986cb}, one obtains
\begin{equation}
\begin{split}
    \tilde S_\text{gf} &\propto\int 
        \tfrac12\tilde h_\mu\wedge\star\Delta \tilde h^\mu
        +\sum_{i={4-d}}^0\bar{\tilde X}^{\sst{i,i-1}}_\mu \wedge\star\Delta\tilde X^{\sst{i}\mu}\\
        &~~~~~~+
    \tfrac12\sum_{i=4-d}^0\sum_{\substack{j=i+1\\i\not\equiv j\pmod2}}^{-i-1}\bar{\tilde X}^{\sst{i,-j}}_\mu\wedge\star\Delta\bar{\tilde X}^{\sst{i,j}\mu}
    ,
    \end{split}
\end{equation}
where we have added \((d-4-i)\)-form antighosts \(\bar{\tilde X}^{(i,j)}_\mu\) of ghost number \(j\)
and used ${\rm{T}}^*M$-valued $p$-form notation
$    \omega_\mu = \frac1{p!}\omega_{\mu\nu_1\dotso\nu_p}\mathrm dx^{\nu_1}\wedge\dotsb\wedge\mathrm dx^{\nu_p}
$. The degree-of-freedom counting is thus as expected:
\begin{equation}
\begin{split}
\tfrac12d(d-3)&=
    \underbrace{d\binom d{d-3}-\binom d{d-2}}_{\tilde h}\\
    &~~~-\sum_{i=4-d}^0(-1)^i\underbrace{(2-i)d\binom d{d-4+i}}_{X^{\sst{i}},\bar X^{\sst{i,j}}}.
    \end{split}
\end{equation}

Now, similar to \eqref{eq:fierz-pauli-extended}, we may  introduce an extra \((d-2)\)-form field \(\tilde B_{\mu\nu_1\dotso\nu_{d-3}}=\tilde B_{[\mu\nu_1\dotso\nu_{d-3}]}\) to write
\begin{equation}
    \tilde H_{\mu\nu_1\dotso\nu_{d-3}} = \tilde h_{\mu\nu_1\dotso\nu_{d-3}} +
    \frac1{\sqrt{d-2}}\tilde B_{\mu\nu_1\dotso\nu_{d-3}},
\end{equation}
such that \(\tilde H_{\mu\nu_1\dotso\nu_{d-3}}\) is an arbitrary \(\mathrm T^*M\)-valued \((d-3)\)-form. Then $
    \tilde S_\text{gf}
    = \frac{1}{\tilde \kappa^2}(S_{d-3}^{\mathrm T^*M} - S_{d-2} + \bar S_{d-3} - S_{d-4})
$, 
where
\begin{subequations}\label{eq:dual-fierz-pauli-extended-detail}
\begin{equation}
\begin{split}\label{eq:dual-fierz-pauli-extended-detail1}
S_{d-3}^{\mathrm T^*M} &= \int \tfrac12\tilde H^\mu\wedge\star\Delta\tilde H_\mu + \sum_{i=4-d}^0\bar{\tilde X}^{\sst{i}\mu}\wedge\star\Delta\tilde X^{\sst{i}}_\mu \\
&\quad\;+\tfrac12\sum_{i=4-d}^0\sum_{\substack{j=i+1\\i\not\equiv j\pmod2}}^{-i-1}\bar{\tilde X}^{\sst{i,-j}}_\mu\wedge\star\Delta\bar{\tilde X}^{\sst{i,j}\mu}
\end{split}
\end{equation}
\begin{equation}\label{eq:dual-fierz-pauli-extended-detail2}
\begin{split} 
    S_{d-2} &= \int \tfrac12\tilde B\wedge\star\Delta\tilde B
    + \sum_{i=3-d}^0\bar c^{\sst{i,i-1}}\wedge\star\Delta c^{\sst{i}}\\
&  ~~~~~~  + \tfrac12\sum_{i=3-d}^0\sum_{\substack{j=i+1\\i\not\equiv j\pmod2}}^{-i-1}\bar c^{\sst{i,-j}}\wedge\star\Delta\bar c^{\sst{i,j}}
\end{split}
\end{equation}
\begin{multline}\label{eq:dual-fierz-pauli-extended-detail3}
    \bar S_{d-3} = \int \sum_{\mathclap{i=3-d}}^0\Big(\bar c^{\sst{i,i-1}}\wedge\star\Delta c^{\sst{i}}   + \tfrac12\sum_{\mathclap{\substack{j=i+1\\i\not\equiv j\pmod2}}}^{\mathclap{-i-1}}\bar c^{\sst{i,-j}}\wedge\star\Delta\bar c^{\sst{i,j}}\Big)\\
   \hfill+\sum_{\mathclap{i=5-d}}^0\Big(\bar\Lambda^{\sst{i,i-1}}\wedge\star\Delta\Lambda^{\sst{i}}+
    \tfrac12\sum_{\mathclap{\substack{j=i+1\\i\not\equiv j\pmod2}}}^{\mathclap{-i-1}}\bar\Lambda^{\sst{i,-j}}\wedge\star\Delta\bar\Lambda^{\sst{i,j}}\Big)\\
    + \tfrac12\tilde\phi\wedge\star\Delta\tilde\phi
\end{multline}
    \begin{equation}\label{eq:dual-fierz-pauli-extended-detail4}
\begin{split} 
    S_{d-4} &= \int\tfrac12\tilde \phi\wedge\star\Delta \tilde \phi\   + \sum_{i=5-d}^0\bar\Lambda^{\sst{i,i-1}}\wedge\star\Delta\Lambda^{\sst{i}}\\
& ~~~~~~+
    \tfrac12\sum_{i=5-d}^0\sum_{\substack{j=i+1\\i\not\equiv j\pmod2}}^{-i-1}\bar\Lambda^{\sst{i,-j}}\wedge\star\Delta\bar\Lambda^{\sst{i,j}},
\end{split}
\end{equation}
\end{subequations}
where we have introduced the  $p$-form BV triangles of (anti-)ghosts $c^{\sst{i}}, c^{\sst{i,j}}, \Lambda^{\sst{i}}, \Lambda^{\sst{i,j}}$ as required. These are, in an obvious sense \footnote{That is, as dual $p$-form and $(d-p-2)$-form theories. On the other hand $B$ and $\tilde B$, for example, should be regarded as dual fields once one recalls that the $(d-2)$-form  is a $\rm{T}^*M$-valued $(d-3)$-form with a symmetry constraint and, so, dual to a $\rm{T}^*M$-valued $1$-form with a symmetry constraint.}, dual to the (anti-)ghosts of \eqref{eq:fierz-pauli-extended}; for instance the \((d-4)\)-form \(\tilde \phi\) and its associated and BV triangle \(\Lambda^{\sst{i}}, \bar\Lambda^{\sst{i,j}}\) is dual to the $2$-form field \(B\) in \eqref{eq:fierz-pauli-extended-detail2}.

 Then we may interpret each  term in \eqref{eq:dual-fierz-pauli-extended-detail} as follows: \(S_{d-3}^{\mathrm T^*M}\) is the action for an unconstrained \(\mathrm T^*M\)-valued \((d-3)\)-form gauge field, dual to  \eqref{eq:fierz-pauli-extended-detail1}; \(S_{d-2}\) is the action for a \((d-2)\)-form gauge field, dual to the  \eqref{eq:fierz-pauli-extended-detail2};  \(\bar S_{d-3}\) is the action for  a pair of wrong statistic \((d-3)\)-form fields dual to \eqref{eq:fierz-pauli-extended-detail3};  \(S_{d-4}\) is the action for a \((d-4)\)-form gauge field, dual to the \(S_2\) in \eqref{eq:fierz-pauli-extended-detail4}. Again,  the total degree-of-freedom count correctly matches  $\tfrac12d(d-3)$.

Putting these contributions together,  the dual partition function \(\tilde Z_\text{grav}\) for the dual graviton may be written as
\begin{equation}\label{eq:dual-partition-function-relation}
    \tilde Z_\text{grav} = \frac{Z_{d-3}^{\mathrm T^*M}}{Z^{M\times \mathbb{R}}_{d-4}(Z^{M\times \mathbb{R}}_{d-3})^2Z^{M\times \mathbb{R}}_{d-2}}.
\end{equation}

\subsection{The duality anomaly for linearized gravity}
Using \eqref{eq:partition-function-relation} and \eqref{eq:dual-partition-function-relation}, the duality anomaly is given by
\begin{equation}
 \frac{Z_\text{grav}}{\tilde Z_\text{grav}}
   = \frac{Z_1^{\mathrm T^*M}}{Z_{d-3}^{\mathrm T^*M}}
    \cdot\frac{Z^{M\times \mathbb{R}}_{d-2}}{Z^{M\times \mathbb{R}}_0}
    \cdot
    \frac{Z^{M\times \mathbb{R}}_{d-4}}{Z^{M\times \mathbb{R}}_2}
    \cdot
    \left(
        \frac{Z^{M\times \mathbb{R}}_{d-3}}{Z^{M\times \mathbb{R}}_1}
    \right)^2.
\end{equation}
Noting each factor is a ratio of dual $E$-valued $p$-form partition functions, upon generalizing  \cite{Donnelly:2016mlc} to vector-bundle-valued \(p\)-form fields, the positive-energy modes yield the Ray--Singer torsion for $E\to M$, while the zero modes and instantons combine into the Reidemeister torsion up to an anomaly given by $(\kappa/\tilde \kappa)^{\frac12\chi(M;\mathrm T^*M)} $ 
to yield \eqref{eq:grav-anomaly}.  

In $d=4$ \footnote{More generally, we can consider any $\frac{d}{2}=p+1$ for arbitrary ${\rm {T}^*M}$-valued $p$-form and $(d-2-p)$-form dual formulations of linearized gravity.}, one can add a gravitational $\theta$-term, the linearization and resolution of $\theta \int R \wedge R$. In this case, the  duality is extend to a modular $\operatorname{SL}(2; \mathbb{Z})$ action on $\tau = \frac\theta2+ \mathrm i\frac{2\pi}{\kappa^2}$. The partition function is then a modular form, $
\tilde Z(\tau)=\mathrm e^{\mathrm i\sigma}\tau^{-\frac{1}{4}(\chi-\sigma)} \bar{\tau}^{-\frac{1}{4}(\chi+\sigma)} Z(-1 / \tau)
$, up to a phase given by the twisted Hirzebruch signature $\sigma(M; {\rm {T}^*M})$. This  is  in direct analogy to  Abelian S-duality  \cite{Witten:1995gf}, with the corresponding  phase   identified in  \cite{Olive:2000yy,Metlitski:2015yqa,Donnelly:2016mlc}. Accordingly, the  duality interchanges the linearized first Bianchi identities \footnote{The dual second Bianchi identities are rotated amongst themselves \cite{deMedeiros:2002qpr}.} and equations of motion of the dual gravitons  \cite{deMedeiros:2002qpr}, which in $d=4$ can be placed into an $\text{SL}(2, \mathbb{Z})$ doublet. Finally, if we dimensionally reduce the dual gravitons to $d=4$, we obtain dual graviphotons (Maxwell gauge potentials) that are related  by an Abelian S-duality.

Arriving at these conclusions  relied on the Cheeger--Müller theorem equating the Ray--Singer and Reidemeister torsions. In doing so,  \(\mathrm T^*M\) must be assumed flat \footnote{That is, $M$ is a flat Riemannian manifold.}, but in establishing the \emph{existence} of  anomalies this is no loss.  Less trivially,  the  \(\mathrm T^*M\)-valued $p$-form instantons must be taken to be in  ${\rm H}^{p+1}(M; \mathbb{Z})$, a matter we turn to  now.

\subsection{The instanton sector}\label{ssec:instantons}
The gravitational duality anomaly \eqref{eq:grav-anomaly}  depends crucially on the contributions from zero modes and instantons. Indeed, if one were to ignore them, the duality anomaly would vanish in \emph{even} dimensions and not in odd dimensions \footnote{The same observation  was made for  \(p\)-forms  in \cite{Donnelly:2016mlc}.}, contradicting  standard expectations.  

The zero modes  straightforwardly correspond to the zero eigenvalues of the Laplace--de~Rham operator for (vector-bundle-valued) differential forms, including those for ghosts.  On the other hand, the prescription for instantons is more subtle. For \(p\)-form electrodynamics, the instanton sectors are given by topologically inequivalent gerbes or, equivalently, the integer cohomology classes \(\operatorname H^{p+1}(M;\mathbb Z)\); the possible gauge fields are connections on one of the possible gerbes on \(M\).
There is, however,  no analogue for  general relativity on a fixed smooth spacetime manifold \(M\); all metrics are  (nondegenerate) sections of the one fixed bundle, \({\rm Sym}^2(\mathrm T^*M)\). Rather, the gravitational partition function includes the sum over all possible topologies and smooth structures of the spacetime manifold \(M\). Apart from being technically  difficult to compute, the  ``linearization'' appropriate for  massless Fierz--Pauli gravity, which involves a \emph{single} fixed background metric, topology, and smooth structure, is not obvious.

To some extent, this is an issue of semantics in \emph{defining} what the quantum theory of the massless Fierz--Pauli model (and its dual) should be. What ought to be summed-over in the path integral? Here,  we have \emph{defined} the  instanton sectors of the dual graviton theories to be
\begin{equation}\label{eq:instanton_sectors_gravity}
    \text{``}\frac{\operatorname H^n(M;\mathrm TM)}{\operatorname H^{n+1}(M)\oplus\operatorname H^n(M)\oplus\operatorname H^n(M)\oplus\operatorname H^{n-1}(M)}\text{''},
\end{equation}
where $n=2$  for the dual graviton and $n=d-2$ for the dual graviton so that they are related via Poincaré duality. 
This expression is enclosed in quotes since it is not a true quotient, but rather a suggestive notation for the prescription that there is a multiplicative factor corresponding to a sum over \(\operatorname H^n(M;\mathrm TM)\), and the inverse of a factor corresponding to a sum over \(\operatorname H^{n+1}(M)\oplus\operatorname H^n(M)\oplus\operatorname H^n(M)\oplus\operatorname H^{n-1}(M)\). The intuition behind \eqref{eq:instanton_sectors_gravity} is that the (dual) graviton (resolved to a $\mathrm T^*M$-valued $(d-3)/1$-form) instantons contribute a factor of \(\operatorname H^n(M;\mathrm TM)\), the cohomology relevant to the integral $\mathrm T^*M$-valued curvatures. However, this  over counts due to the ghost integral curvatures, hence the denominator $\operatorname H^{n+1}(M)\oplus\operatorname H^n(M)\oplus\operatorname H^n(M)\oplus\operatorname H^{n-1}(M)$, which subtracts the ghost instanton contributions.

 Intrinsically, the prescription \eqref{eq:instanton_sectors_gravity} makes sense if the Fierz--Pauli model is considered as a  gauge theory in its own right: we are simply summing over all ways in which the field \(H_{\mu\nu}\) can have nontrivial Čech cocycles, modulo the corresponding nontrivial Čech cocycles for \(B_{\mu\nu}=H_{[\mu\nu]}\) and the corresponding (anti)ghosts, and  similarly  for the dual  graviton. This seems a natural, and almost inevitable, contribution to the path integral. 
Extrinsically, one can always combine a massive Fierz--Pauli model with its dual to manifest a  classical \(\operatorname U(1)\)  duality symmetry that should only be   anomalous  in  even  dimensions, since anomalies are given by certain characteristic classes of even degree. This requires the instanton sectors be given by  \eqref{eq:instanton_sectors_gravity}. Turing this around, duality anomaly freedom (or modularity) can be used as a heuristic identifying the correct path integral, which cannot be inferred from the classical action alone. 

\section{Discussion}\label{sec:discussion}

Providing an instanton prescription and resolving $\text{Sym}^2({\rm T}^* M)$ into $\Omega^1(M, {\rm T}^* M)$, we have shown that the linear graviton duality anomaly is controlled by the Euler characteristic and, when a $\theta$-term is included, that the partition function is a modular form on $\frac\theta2+ i\frac{2\pi}{\kappa^2}$. A number of implications and generalizations present themselves.

Most obviously, the resolution method should be directly applicable to the numerous exotic dualities for  various spins \cite{Hull:2001iu,deMedeiros:2002qpr},  sufficing to compute the duality anomalies and to identify an instanton sector prescription.

More ambitiously, recall  the invocation of the Cheeger--Müller theorem required  \(\mathrm T^*M\) be flat so as to form a representation of \(\pi_1(M)\). Physically there is no reason for this restriction: the anomaly should always exist and be an invariant of the smooth structure of \(M\). This suggests an extension of the Cheeger--Müller theorem that applies for non-flat vector bundles. On the ``analytic'' side, it would  involve the zeta-regularized determinant and zero modes of the Lichnerowicz operator  differing from the Laplace--de~Rham operator \(\Delta\) by curvature terms,
$
    \Delta_{\mathrm{L}} h_{\mu \nu}=\Delta h_{\mu \nu}-2 R_{\mu \rho \nu \sigma} h^{\rho \sigma}+2 R_{(\mu}{ }^\rho h_{\nu) \rho},
$
for which there exist known results \cite{Christensen:1979iy}. Pushing this even further, one could consider the sum  over background metrics, then differentiable structures and even topologies, but here one would expect the anomaly  to factorize over the sum.   

Of course, here we have only treated the linearized graviton partition functions. Ultimately,  the full interacting theories should be considered.  However, the linearized  duality anomaly calculation is nonetheless an important step since either it persists in the full theory (which is perhaps the default expectation) or it is miraculously canceled upon including interaction (in which case knowing the linearized anomaly independent is an essential ingredient). 

Finally,  dual gravitons, and related generalizations thereof,  arise in various approaches to quantum gravity \cite{Borissova:2023yxs}, M/E-theory  \cite{Hull:2000zn, Hull:2000ih, Hull:2000rr, West:2001as, Henneaux:2016opm, Hohm:2018qhd, Tumanov:2017whf, Glennon:2020uov, Hull:2022vlv,Boulanger:2024lwk} and generalized symmetries \cite{Hinterbichler:2022agn,Benedetti:2021lxj,Benedetti:2023ipt,Benedetti:2022zbb,Gomez-Fayren:2023qly,Hull:2024bcl,Hull:2024ism};  the duality anomaly may have non-trivial consequences in such contexts. For instance, building on the argument of \cite{Borsten:2021pte}, the vanishing of the gravitational and Abelian $3$-form duality anomalies in M-theory (since $d=11$), implies  anomaly freedom for type IIA string theory. The anomaly of the IIA massless sector is canceled precisely by that of the M-theory Kaluza--Klein tower \footnote{The details will be given in \cite{Borsten:2025aaa}.}. Put another way, insisting on duality anomaly freedom in type IIA implies the existence of the M-theory Kaluza--Klein spectrum. More radically, it is tempting to speculate that, just as the S-duality of $d=4, \mathcal{N}=4$ super Yang--Mills theory is a consequence of $d=6$ string/string duality \cite{Duff:1994zt} and the \emph{self-dual} $d=6, \mathcal{N}=(2,0)$ theory \cite{Witten:2009at}, the modularity of the $d=4$ linearized graviton partition function observed here is a remnant  of the conjectured \emph{self-dual}   $d=6, \mathcal{N}=(4,0)$  ``gravi-gerbe'' theory \cite{Hull:2000rr,Borsten:2017jpt, Hull:2022vlv}.

\begin{acknowledgments}

MJD is supported in part by the STFC Consolidated Grant ST/X000575/1.

\end{acknowledgments}

\end{document}